\documentclass[a4paper,conference]{IEEEtran}
\usepackage{amssymb}
\usepackage{cite}
\usepackage{graphicx}
\usepackage{psfrag}
\usepackage{subfigure}
\usepackage{url}
\usepackage{amsmath}
\usepackage{mathrsfs}
\usepackage{color}
\newcommand{\prob}[1]{\mathsf{Pr}\left( #1 \right)}

\newcommand{\remove}[1]{}

\newcommand{\be}{\begin{equation}}
\newcommand{\ee}{\end{equation}}

\newcommand{\EMone}{$\mathsf{EM}_1$}
\newcommand{\EMtwo}{$\mathsf{EM}_2$}

\newtheorem{theorem}{Theorem}
\newtheorem{lemma}{Lemma}
\newtheorem{remark}{Remark}

\begin{document}
%%%%%%%%%%%%%%%%%%%%%%%%%%%%%%%%TITLE%%%%%%%%%%%%%%%%%%%%%%%%%%%%%%%%%%%%%%%%
\title{On Distributed Computation in Noisy Random Planar Networks}

\author{ \authorblockN{Yashodhan Kanoria and D.~Manjunath}
  \authorblockA{Department of Electrical Engineering \\
    Indian Institute of Technology  Bombay \\
    Mumbai 400076  INDIA \\
    \texttt{yashodhan,dmanju@ee.iitb.ac.in}} }

% make the title area
\maketitle

%%%%%%%%%%%%%%%%%%%%%%%%%%%%%%ABSTRACT%%%%%%%%%%%%%%%%%%%%%%%%%%%%%%%%%%%%%%%
\begin{abstract}
  THIS PAPER IS ELIGIBLE FOR THE STUDENT PAPER AWARD. We consider
  distributed computation of functions of distributed data in random
  planar networks with noisy wireless links. We present a new
  algorithm for computation of the maximum value which is order
  optimal in the number of transmissions and computation time. We also
  adapt the histogram computation algorithm of Ying et al
  \cite{Ying06} to make the histogram computation time optimal.
\end{abstract}


\section{Introduction}
We consider distributed or `in-network' computation of functions of
sensing data in sensor networks in two dimensions.  The sensor nodes
collect sensing data and communicate with other nodes in a limited
range over noisy wireless links. Our interest is in efficient
evaluation of specific functions of sensing data. Latency, energy cost
and throughput are the efficiency measures. We assume that
communication costs dominate the time and energy cost for the
computations.

Two types of wireless networks are considered in the literature
studying function computation---random planar networks
\cite{Giridhar05,Khude05,Ying06} and collocated or broadcast networks
\cite{Gallager88,Newman04,Giridhar05,Goyal05}. Results for collocated
networks are useful in solving corresponding problems for random
planar networks. While \cite{Giridhar05,Khude05} consider noise-free
links, \cite{Gallager88,Newman04,Goyal05,Ying06} consider noisy links.
Computation of the histogram, and hence that of any symmetric
function, in a random planar network with noisy links is considered in
\cite{Ying06}.  A protocol requiring $\Theta\left(n\log \log n\right)$
transmissions to compute the histogram under the strong noise model is
described. The lower bound is open. \cite{Ying06} is the only study of
distributed computation in a noisy random planar network and it does
not obtain the computation time (which is not the same as the energy
for a random planar network).  In this paper, we show how to
effectively use the coding theorem of \cite{Rajagopalan94} to reduce
the computation time in noisy random planar networks for maximum (MAX)
and histogram functions. Also, our protocol for MAX requires
$\Theta\left(n\right) $ transmissions.

\section{Model and Problem Description}
Sensor nodes $1,2, \ldots, n$ are uniformly distributed in $[0,1]^2.$
The transmission range of all the nodes is $r_n.$ We use the protocol
model of interference \cite{Gupta00}.  Let $\rho_{i,j}$ be the
distance between nodes $i$ and $j$. A transmission of node $i$ can be
received at node $j$ if $\rho _{i,j} \le r_n$, and for every other
simultaneously transmitting node $k$, $\rho_{k,j} \ge \left(1+\Delta
\right)r_n$, for some constant $\Delta$.  If $\rho_{i,j} < r_n$ and
there exists a node $k$ that is transmitting at the same time as node
$i$ such that $\rho_{k,j} \leq \left(1+\Delta \right)r_n$, then there
is a collision at node $j.$

Time is slotted and all nodes are synchronized to these slots. The
slot duration is equal to the bit duration. At time $t$, sensor node
$i$ has data $x_i(t)$ with $x_i(t)$ taking values from a finite set.
For simplicity, we assume $x_i(t)$ is $\{0,1\}$-valued.  (A general
finite set only changes the results by a constant factor.)  Define
$x(t):=\left[ x_1(t), x_2(t), \ldots, x_n(t) \right].$ Our interest is
in efficiently computing $f(t):=\phi(x(t))$ at some instants of time.
$f(t)$ is to be made available at a single sink node $s$, which is a
node close to the center of the unit square. In general, $f(t)$ may be
required at a subset of the nodes.

There are many ways in which the computation of $f(t)$ for different
$t$ can be scheduled. Three such computation models are usually
discussed in the literature. (1) \emph{One shot} computation is a
one-time computation of $f(\cdot)$ and, without loss of generality, we
consider only $f(0).$ (2) In \emph{pipelined} computation $f(t)$ is to
be computed for $\{t_k\}_{k=1,2,\ldots}$ with the computations for the
different $t_k$ being pipelined at each of the nodes. (3) In \emph{
  block computation} $x(t)$ is collected for a block of, say, $K$
instants, i.e., for $\{t_k\}_{k=1,\ldots,K}$ and $f(t)$ is computed
for this block of time.

The wireless links in the network are assumed to be binary symmetric
channels with errors independent across receivers. We assume that the
error probability is upper bounded by some $\epsilon_0 < 0.5$.  Thus,
our protocols work for the strong \emph{clairvoyant} adversary model
\cite{Feige00}. Note that the protocols in
\cite{Gallager88,Newman04,Ying06} also work for this noise model.  A
weaker noise model where each bit is in error with probability exactly
$\epsilon_0$ has also been considered in the literature, e.g.,
\cite{Kushilevitz98}.

Different distributed scheduling strategies are possible. In a
collision free strategy (CFS), transmissions are scheduled such that
in the absence of noise, there is no collision at the intended
receivers. Also, information cannot be communicated by a node by
avoiding a transmission in its scheduled time slot. This means that
even though our model for the noisy network does not allow bit
erasures, the protocols designed are robust in their presence.
Erasures, if they occur, can be immediately identified as errors for
such protocols.  A CFS may be \emph{oblivious} or
\emph{non-oblivious}.  In an \emph{oblivious} protocol, the
transmission schedule is fixed beforehand. In a \emph{non-oblivious}
protocol, the data values transmitted in previous slots determine the
evolution of the schedule.  A non-oblivious CFS designed for a
noiseless network can result in collisions when used in a noisy
network. An oblivious protocol ensures that such collisions do not
occur. We look for oblivious protocols.

Let $E_t$ be the energy consumed for transmission of one bit, $E_r$ be
the energy consumption at the receiver of a transmission. We ignore
processing energy. In general, $E_t$ depends on $r_n$, and hence on
$n$. However, results in \cite{Khude05} indicate that for sufficiently
large $n$ ($\ge 5000$), it is reasonable to assume that the
transmission energy is constant and independent of $n$. We follow this
assumption. Note that this is different from the energy model of
\cite{Ying06}. If we let $E_r < < E_t$, the number of transmissions
required in the protocol can be used as a proxy for the energy
consumed. Energy model \EMone will refer to the case when $E_r$ is not
to be ignored and \EMtwo will refer to the case when it can be
ignored. In \EMtwo, we simply count the number of transmissions.
Multiplying the energy in \EMtwo by $k_5\left( \sqrt{\frac{\log n}{n}}
\right)^\alpha$ for some constant $k_5$ and $\alpha$ gives the energy
in the model of \cite{Ying06}.

As was mentioned earlier, we will consider the evaluation of the
maximum (MAX) and histogram of $x(t).$ Note that the MAX is the same
as the OR function for binary data.

\subsection{Summary of Results and Organization of the Paper}   
\label{sec:summary}
We describe a single oblivious protocol for computation of MAX in a
noisy network that matches the lower bounds for both time and
transmissions of the optimal CFS in a noise free network. This may be
compared against the $\log \log n$ penalty in number of transmissions
that seems inevitable for the histogram function \cite{Ying06}.  We
also show how to achieve the trivial lower bound in time for histogram
computation in a noisy random planar network, using a suitable
modification of the algorithm presented in \cite{Ying06}.  The number
of transmissions is the same as that in \cite{Ying06}, i.e.,
$\Theta\left(n\log \log n\right)$.

\section{MAX in a Noisy Random Planar Network}
\label{section:main}
Like in \cite{Khude05}, we wish to evaluate $f=\textup{MAX/OR}$.
Following \cite{Giridhar05,Khude05,Ying06}, we propose a two-stage
algorithm which is implemented by dividing the operational area into
appropriately sized cells, performing intra-cell computation, or
fusion, and then propagating the fused results towards the sink with
appropriate combination on the way.

\begin{figure}[tb]
  \centering
  \includegraphics[width=1.6in,keepaspectratio]{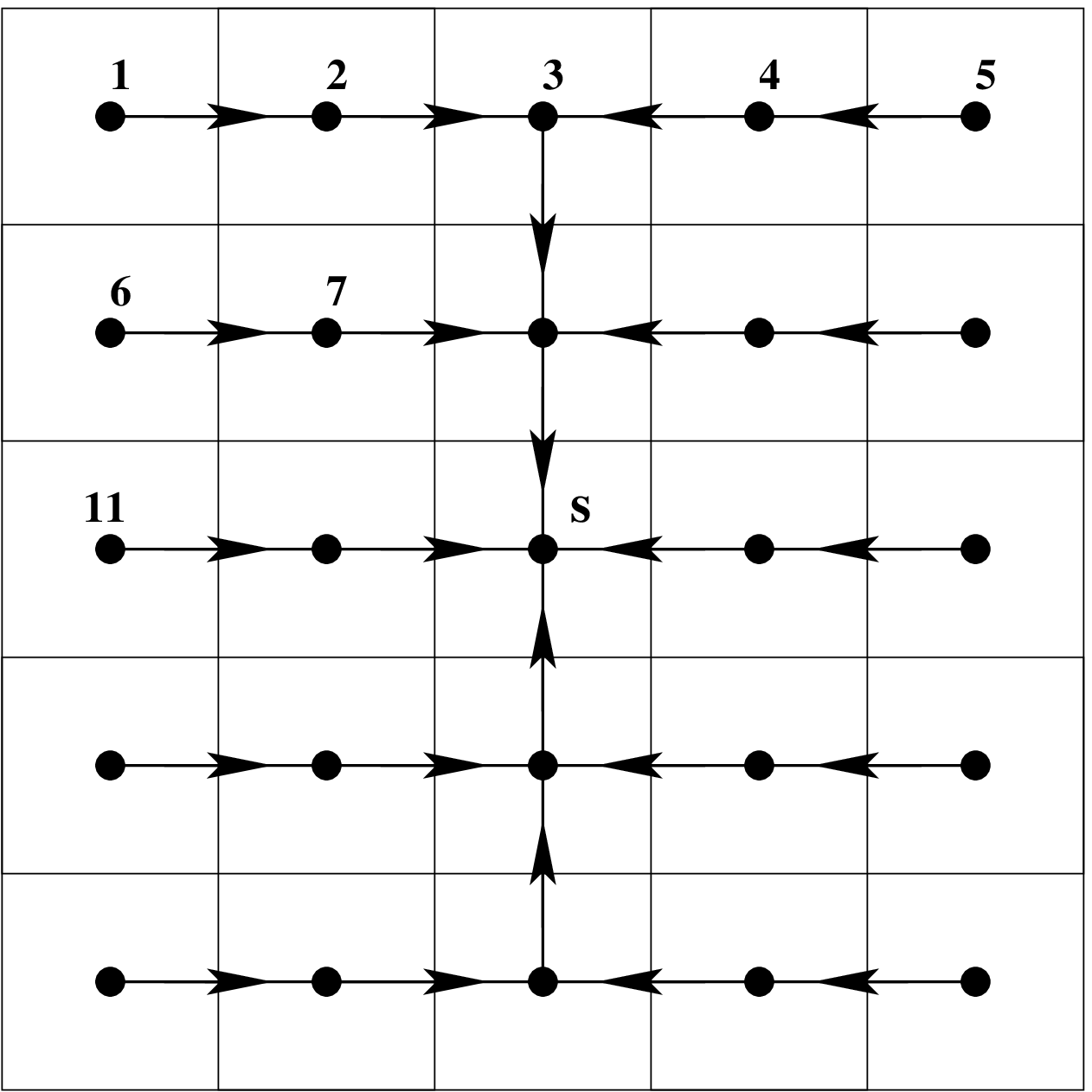}
  \caption{The tessellation of the unit square with labeling of
    cells. The spanning tree used in inter-cell communication is also
    shown. }
  \label{fig:sptree}
\end{figure}

Following \cite{Xue04}, we perform a tessellation of the unit square
into square cells of side $l= \left\lceil \sqrt{\frac{n}{2.75 \log n}}
\right\rceil^{-1}.$ This gives us a total of $M_n=\left\lceil
  \sqrt{\frac{n}{2.75 \log n}} \right\rceil^2.$ Label the cells as
$S_j, j=1,2,\ldots,M_n$, from left to right and from top to bottom as
shown in Fig.~\ref{fig:sptree}.  Let $N_j$ denote the number of sensor
nodes in cell $S_j.$ Of course $\sum_{j=1}^{M_n} N_j =n.$ Choose
$r_n=\sqrt{\frac{13.75 \log n}{n}} \approx \sqrt{5} l.$ Since $l \le
\frac{r_n}{\sqrt 2}$ the nodes in each of the cells form a single hop
network.  Lemma~3.1 of \cite{Xue04} provides us with useful bounds on
the number of nodes in each cell.
%In fact, we have chosen $K=2.75$ for this purpose.

\begin{lemma}
  For any $\mu > \mu^* =0.9669$,
  \begin{eqnarray}
    \lim_{n\rightarrow \infty} \prob{\max_j |{N}_{j}  - 2.75
      \log n|\le 2.75 \mu \log n} = 1 
    \label{eq:nodebound2-ij}
  \end{eqnarray}
\end{lemma}

The above lemma means that for $j=1,2,\ldots,M_n,$ with probability
$1$, as $n\rightarrow \infty$
\begin{eqnarray}
  \label{eq:nodebound}
  && 0.091 \log n  \ \le \  {N}_{j} \ \le \ 5.41 \log n  
  \label{eq:nodebound-ij}
\end{eqnarray}

We form a spanning tree of the cells rooted at cell $S_s$ with edges
along the central vertical axis and along all horizontal rows of cells
as shown in Fig.~\ref{fig:sptree}. Let $\pi (j)$ denote the index of
the parent of cell $j$, i.e. cell $S_{\pi (j)}$ is the parent of
$S_j$, $j=1,2,\ldots,s-1,s+1,\ldots,M_n$.  We arbitrarily select one
node in each cell $S_j$ to be the \textit{cell center} $C_j$ (except
that the sink node is made the cell center of $S_s$). This is possible
with high probability since (\ref{eq:nodebound}) tells us that each
cell is occupied.  The cell centers of two adjacent cells are one-hop
neighbors, since $l \le \frac{r_n}{\sqrt 5}$.

\subsection{Stage 1: Intra-cell Computation}
At the end of the first stage, the maximum of the values in each cell
$S_j$ is to be made available to $C_j.$ Since each cell behaves like a
single-hop network, we can use a noisy broadcast network protocol for
this purpose.

It helps to review the protocol for a noiseless random planar network
\cite{Khude05}.  Each node transmits its bit to the cell center. From
(\ref{eq:nodebound}), it is sufficient to give each cell $\lfloor
5.41\log n \rfloor$ time slots.  During the computation in each cell,
at most $k_1 := \left(2\lceil \frac{ \left(1+\Delta \right) r_n}{l}
  \rceil +1\right)^2-1$, i.e., a constant number of adjacent cells
will be disabled due to interference. Thus, with high probability,
this stage will require at most $(k_1+1)\lfloor 5.41 \log n \rfloor$
slots, i.e., $\Theta \left (\log n\right) $ time using $\Theta \left
  (n\right) $ total transmissions.
%NEED TO MAKE THIS EXACT.

For a noisy network, \cite{Ying06} describes a protocol using which
the aggregation in a cell can be performed in $\Theta \left (\log n
  \log \log n\right) $ slots (with scheduling of cells as above) using
$\Theta \left (n\log \log n\right) $ total transmissions with overall
probability of error $O\left(\frac{1}{\log n}\right) $. Using this
intra-cell protocol, we can collect the histogram of the cell sensor
readings at the cell-center.  However, this is more information than
we need.  Instead, we adapt the algorithm of \cite{Newman04} to reduce
the complexity of the intra-cell protocol to $O\left(n\right)$
transmissions. We explain this adaptation next.

Define the `witness' for cell $S_j,$ $\hat{W}_j,$ as follows.
$\hat{W}_j=i_0$ if $i_0$ is the minimum $i$ for which $x_i=1$ and $i
\in S_j.$ If $x_i=0$ for all $i \in S_j$, an arbitrary node can be
designated as $\hat{W}_j.$

%Describe Newman in the Appendix; 
%Obtain $k_2$  and also derive the correctness probability. 

Each cell is a single-hop network and we run the `witness discovery'
protocol described in the proof of Theorem~2 of \cite{Newman04}. From
Theorem~2 of \cite{Newman04}, witness discovery in $S_j$ can be
achieved by an oblivious protocol with less than $k_2 N_j$
bit-transmissions in $S_j$, for some constant $k_2$.  At the end of
the execution of this protocol, $C_j$ identifies some $W_j$ as the
witness for $S_j$ such that $\prob{W_j\neq \hat{W}_j}<\epsilon_1$ for
any desired constant $\epsilon_1 >0$.  We can therefore achieve
$\prob{x_{W_j} \neq \max_{i; i \in S_j} \{x_i\} } < \epsilon_1,$ for
any $\epsilon_1>0$ i.e., $W_j$ has the correct value of the maximum of
the values in cell $j$ with error probability that can be upper
bounded by any constant.  If $\epsilon$ is the desired bound for the
error probability in computing $f,$ we choose
$\epsilon_1=\frac{\epsilon}{2}$ for each cell.

From Observation~2.1 in \cite{Newman04}, $C_j$ can distribute the
identity of $W_j$ to all nodes in $S_j$ using $\Theta \left (\log
  n\right) $ transmissions to ensure that the identity of $W_j$ is
known to all nodes in the cell with error probability
$O\left(\frac{1}{n}\right)$.  Finally, $W_j$ transmits $x_{W_j}$,
$\Theta \left (\log n\right) $ times to $C_j$ to ensure that the
probability of $C_j$ having the wrong value of $x_{W_j}$ is
$O\left(\frac{1}{n}\right)$.  Let $f_j$ denote the value of $x_{W_j}$
decoded by $C_j$ from these transmissions. Note the use of $\Theta
\left (\log n\right) $ transmissions and not $\Theta \left (\log
  N_j\right) $ transmissions in the last two steps to ensure error
probability is $O\left(\frac{1}{n}\right)$ for each of the cells.

Thus, $\Theta \left (\log n\right) + k_2 N_j$ transmissions suffice to
complete Stage~1 in cell $S_j$. Using (\ref{eq:nodebound}) we can find
constants $k_3$ and $k_4$ such that Stage~1 requires $a_j N_j = \Theta
\left (\log n\right) + k_2 N_j $ transmissions in $S_j$ with $k_3 <
a_j < k_4 \ \forall \ j=1,2,\ldots,M_n$. This, along with the bound on
the number of interfering neighbors for a cell, means that Stage~1 can
be completed in the entire network in $\Theta \left (\log n\right) $
time using $\Theta \left (n\right) $ transmissions.

While the probability of error in correctly identifying $W_j$ in each
cell $j$ is upper bounded by a constant, we do not yet know a
`network-wide' error probability. But then how do we define the error
in the network? An obvious definition of a `network-error' is the
event of there being an error in identifying the witness in one or
more cells.  A little consideration shows that a more appropriate
definition of the `error' would be as follows. Let $f:=\max_{1 \leq i
  \leq n} \{x_i \}$ be the true value of the MAX function, and let
$\hat{f}:=\max_{1 \leq j \leq M_n} \{f_j \}.$ We will say that
$\gamma_1,$ a Stage~1 error, has occurred, if $f\neq \hat{f}.$ We show
that the probability of $\gamma_1$ is bounded by a constant for large
$n$.

An error in stage $1$ (as defined above) occurs in two ways,
\begin{enumerate}
\item $f=1$ and $f_j=0 \ \forall \ j$. In this case, there must be a
  cell $S_{\hat{j}}$ that has at least one $1$.  Now the probability
  of missing all $x_i=1$ for $i \in S_{\hat{j}}$ in identifying the
  witness for cell $S_{\hat{j}}$ is bounded by $\frac{\epsilon}{2}$.
  The probability of finding a witness $W_{\hat{j}}$ with
  $x_{W_{\hat{j}}}=1$ but having $f_{\hat{j}}=0$ is
  $O\left(\frac{1}{n}\right)$. Thus the probability of having
  $f_{\hat{j}}=0$ is bounded by $\frac{\epsilon}{2} +
  O\left(\frac{1}{n}\right)$.  This gives us a bound of
  $\frac{\epsilon}{2} + O\left(\frac{1}{n}\right)$ on the probability
  of the event $\gamma_1.$
\item $f=0$ and $f_j=1$ for one or more $j.$ Since $f=0,$ we have
  $x_{W_j}=0 \ \forall \ j.$ The probability of obtaining the wrong
  value of $f_j$ is $O\left(\frac{1}{n}\right)$ for $1 \leq j \leq
  M_n.$ Since there $M_n= \Theta \left (\frac{n}{\log n}\right) $
  cells, from the union bound, the probability of $\gamma_1$ is
  $O\left(\frac{1}{\log n}\right).$
\end{enumerate}
Thus the error probability for the Stage~1 is bounded by
$\frac{\epsilon}{2}+O\left(\frac{1}{\log n}\right)$.

Using the proof technique of Theorem~4 of \cite{Newman04}, an
\textit{oblivious} protocol can be designed to achieve the same
asymptotic time and transmissions complexity as the query based
protocol. The initial phase in which each node transmits its data
value a constant number of times and is heard by all other nodes in
its cell requires $\Theta \left (n \log n\right) $ energy under
\EMone. We already know that the energy is $O\left(n \log n\right)$
under \EMone, since each transmission can be heard by at most $\Theta
\left (\log n\right) $ nodes. Hence, the total energy consumption of
the first stage is $\Theta \left (n \log n\right) $ under \EMone.
Because of the phase in which the identity of $W_j$ is distributed,
this is also the energy consumption for the non-oblivious version
described earlier.
\begin{lemma}
  Stage~1 can be completed by an oblivious protocol in time $\Theta
  \left (\log n\right) $ using $\Theta \left (n\right) $ transmissions
  (same as energy under \EMtwo) with energy consumption $\Theta \left
    (n\log n\right) $ under \EMone, such that the overall probability
  of error is bounded by $\frac{\epsilon}{2}+O\left(\frac{1}{\log
      n}\right)$.
\label{lemma:stage-1}
\end{lemma}

\subsection{Stage 2: Inter-cell Computation}
\label{subsection:intercell}
As in \cite{Giridhar05,Khude05,Ying06}, in Stage~2, data is passed
along the tree towards the sink, first horizontally and then
vertically between the cell centers $C_j$ (see Fig.~\ref{fig:sptree}).
Let $\hat{f}_j = \max_{{j^\prime}}\{ {f_{j^\prime}} \}$ where
$S_{j^\prime}$ are the cells in the subtree rooted at cell $S_j$,
excluding cell $S_j$. $\max\{ f_j, \hat{f}_j\}$ is passed to the cell
center of the parent cell $C_{\pi (j)}$. Data thus moves along the
spanning tree towards the sink.

In this stage, parallel communication of data can reduce the time
complexity.  Recall that when $C_j$ transmits, at most $k_1$ other
adjacent cell centers will not be able to receive other transmissions
in the same slot. Note that we need bidirectional flow of data along
the edges of the tree to use \cite{Rajagopalan94}.  Since the vertices
of the tree have a maximum degree of $d=4$, for an active link, at
most $k_5=4(k_1+1)$ links incident on the cells in the neighborhood
will be disabled. Thus, all links can be activated at least once each
in at most $k_5$ slots. We can therefore consider each link in the
spanning tree to be a $1$-bit per `logical slot' link with each
logical slot corresponding to $k_5$ real slots.  Hereafter, we
consider only logical time slots.

It is easy to see that the coding theorem of \cite{Rajagopalan94} is
applicable to the logical network defined above.

\begin{theorem}
  \label{thm:Rajagopalan}
  (Theorem 1.1 of \cite{Rajagopalan94}) Any protocol which runs in
  time $T$ on an $N$-processor network of degree $d$ having noiseless
  communication channels, can, if the channels are noisy (each a
  binary symmetric channel of capacity $C$, $0 < C \le 1$), be
  simulated on that network in time $O\left(T\frac{1}{C} \log (d+1) +
    \frac{1}{C} \log N\right)$ with probability of error $e^{-\Omega
    \left (T\right)}$.
\end{theorem}
\begin{remark}
  The protocol described in \cite{Rajagopalan94} to achieve this uses
  \textit{tree codes.} \cite{Peczarski06} gives the best available
  construction of tree codes---a randomized construction that works
  correctly with probability arbitrarily close to 1.  The computation
  required for maximum likelihood decoding used in the proof of
  Theorem~\ref{thm:Rajagopalan} grows exponentially with the depth of
  the tree and an efficient method for decoding still needs to be
  found to make it computationally tractable.  We ignore computational
  expense in this paper.
\end{remark}

The depth of the tree is $\Theta \left (\sqrt{\frac{n}{\log n}}\right)
$ and there are $\Theta\left(\frac{n}{\log n} \right)$ nodes in the
tree.  Hence Stage~2 would require $\Theta\left(\sqrt{\frac{n}{\log
      n}}\right)$ time slots and $\Theta\left(\frac{n}{\log n}\right)$
transmissions in a noiseless network. We can modify this protocol as
follows to make it possible to use Theorem~\ref{thm:Rajagopalan} to
simulate it in a noisy network---for each time slot in which a link is
not activated send a dummy bit on the link (say $0$). Now, from
Theorem~\ref{thm:Rajagopalan}, this protocol can be simulated in the
noisy network with only constant factor increase in time because we
have $d=4$. However, a naive implementation of the protocol of
\cite{Rajagopalan94} would need $\Theta \left(\left(\frac{n}{\log
      n}\right)^{\frac{3}{2}}\right)$ transmissions, compared to
$\Theta \left (\frac{n}{\log n}\right) $ transmissions for the
noiseless network.  There is an increase of $\Theta \left(
  \sqrt{\frac{n}{\log n}} \right)$, i.e., $\Theta (\frac{1}{r_n}),$
because the protocol of \cite{Rajagopalan94} requires every $C_j$ to
transmit in every slot, while in the noiseless network, each would
transmit just once.

We can reduce the number of transmissions by securing the
transmissions at each level of the tree by repetition coding, i.e.,
transmitting $\Theta \left (\log n\right) $ times on each link. This
increases the number of transmissions to $\Theta \left (n\right) $, an
increase of only $\Theta \left (\log n\right) $ from the noiseless
case. Since each node has to collect the data from the tree below it,
we can use information theoretic arguments to show that even if we
interleave transmissions at different depths of the tree in time, we
cannot do better than this in terms of number of transmissions.  Note
that the overall number of transmissions for the maximum computation
is now just $\Theta \left (n\right) $ i.e., it is optimal.  However,
there is a penalty of $\Theta \left (\log n\right) $ in time, which is
inflated to $\Theta \left (\sqrt{n\log n}\right) $.

Ideally, we would like a way to complete the Stage~2 computation in
$O\left(n\right)$ transmissions (at most as many transmissions as
Stage~1) and in $\Theta \left (\sqrt{\frac{n}{\log n}}\right) $ time
i.e., we would like to not have a penalty either in time or in number
of transmissions.  Surprisingly, this is possible.

\begin{figure}[tb]
  \centering \scalebox{.9}{\input{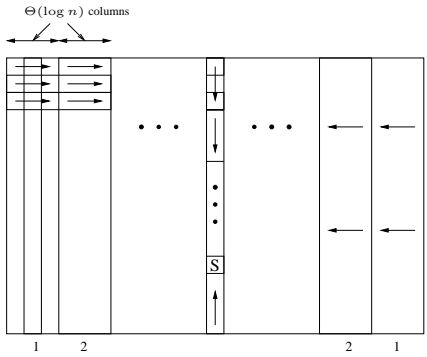}}
  \caption{Sub-stages in the Inter-cell protocol. Part of the unit
    square is shown. $S$ denotes the cell containing the sink.}
	\label{fig:substages}
\end{figure}

This can be achieved by propagation of bits up the tree in a series of
sub-stages, moving $\Theta \left (\log n\right) $ levels (ignoring
edge effects) in $\Theta \left (\log n\right) $ time in each
sub-stage. The first set of sub-stages bring data to the central axis,
and the second set take it to the sink. There are $\Theta \left
  (\sqrt{\frac{n}{(\log n)^{3}}}\right) $ sub-stages.  See
Fig.~\ref{fig:substages}. $1$ and $2$ in the figure correspond to
sub-stages.

At each sub-stage we have $O\left(\frac{1}{r_n}\right)$ linear array
networks with $\Theta \left (\log n\right) $ nodes each. From
Theorem~\ref{thm:Rajagopalan}, we can ensure that the error
probability is $O\left(\frac{1}{n}\right)$ for each linear array.
This gives a bound of $O\left(\sqrt{\frac{1}{n\log n}}\right)$ for the
sub-stage error probability. We have an overall union bound of
$O\left(\frac{1}{(\log n)^2}\right)$ for the error probability of
Stage~2. Note that a vanishing error probability would be impossible
if each sub-stage involved $o\left(\log n \right)$ levels of the tree
in an attempt to reduce the number of transmissions without increasing
the time.  Clearly, each of the $\Theta\left(\frac{n}{\log n} \right)$
links is activated $\Theta \left (\log n\right) $ times in this scheme
i.e., $\Theta \left (n\right) $ transmissions are required in total.
Each transmission is to be heard by a single receiver, hence the
energy under both \EMone and \EMtwo is $\Theta \left (n\right) $.

\begin{lemma}
  Stage~2 can be completed by an oblivious protocol in $\Theta \left
    (\sqrt{\frac{n}{\log n}}\right) $ time using $\Theta \left
    (n\right) $ transmissions (same as energy under \EMtwo) with
  energy consumption $\Theta \left (n\right) $ under \EMone, such that
  the overall probability of error is bounded by
  $O\left(\frac{1}{(\log n)^2}\right)$.
\label{lemma:stage-2}
\end{lemma}

The total error for the computation is
$\frac{\epsilon}{2}+O\left(\frac{1}{\log
    n}\right)+O\left(\frac{1}{(\log
    n)^2}\right)=\frac{\epsilon}{2}+O\left(\frac{1}{\log n}\right) <
\epsilon$ for large $n$.

Combining lemmas~\ref{lemma:stage-1} and \ref{lemma:stage-2}:

\begin{theorem}
  \label{thm:main-theorem}
  In a random planar network with binary data, with binary symmetric
  channels and errors independent across receivers with error
  probability bounded by some $\epsilon_0 < \frac{1}{2}$, with high
  probability the MAX (or $OR$) can be computed with probability of
  error less than any $\epsilon > 0$ for large $n$ using $\Theta \left
    (n\right) $ total transmissions with transmission radius $\Theta
  \left(\sqrt{\frac{\log n}{n}}\right)$ and in a time $\Theta \left
    (\sqrt{\frac{n}{\log n}}\right) $.
\end{theorem}
This matches the corresponding results for noiseless networks in
\cite{Khude05}. The total energy for the protocol described is
$\Theta(n\log n)$ under \EMone and $\Theta(n)$ under \EMtwo.

\section{Discussion}
\label{section:disc}
\subsection{Pipelining}
It is easy to see that a pipelined throughput of $\Theta \left
  (\frac{1}{\log n}\right) $ with a delay of $\Theta \left
  (\sqrt{\frac{n}{\log n}}\right) $ in reaching the sink is also
achievable in the noisy case with a scheme similar to the one given in
\cite{Khude05} for the noiseless case. We alternately perform
iterations of Stage~1 and Stage~2, spending $\Theta \left (\log
  n\right) $ time on each. While doing Stage~2, we allow for one
sub-stage of each of the ongoing computations to be completed. Note
that different sub-stages of different computations can essentially be
done in parallel.

\subsection{Distribution of computation result}
\label{subsection:dist}
The result of the MAX computation can be distributed to all nodes with
only a constant factor overhead in time and number of transmissions.
It can be distributed to the cell centers along the same spanning tree
in a series of sub-stages as in Stage~2
(Section~\ref{subsection:intercell}). Thereafter, each cell center can
broadcast the value $\Theta \left (\log n\right) $ times so that all
nodes have the correct value of the function with error probability
$O\left(\frac{1}{n}\right)$.

\subsection{Non-binary data}
All results can easily be generalized to the case in which sensor data
belongs to any finite set.  Suppose $x_i \in
\mathcal{X}=\{1,2,\ldots,|\mathcal{X}|\} \ \forall \ i=1,2,\ldots,n$.
We can complete the computation of MAX in $\log_2 |\mathcal{X}|$
stages. Each stage consists of two parts. Briefly, the first part is
as described in Section~\ref{section:main}, and reduces the number of
possibilities for the result by a factor of $2$. The second part
involves distribution of the result of the first part as in
Section~\ref{subsection:dist}. Thus the computation can be completed
using $\Theta \left (n \log|\mathcal{X}|\right) $ total transmissions
in a time $\Theta \left (\sqrt{\frac{n}{\log
      n}}\log|\mathcal{X}|\right) $.

\subsection{Block Computation}
The results of Giridhar and Kumar \cite{Giridhar05} carry over
trivially to the case of noisy networks. This is because in block
computation, Gallager's coding theorem can be used to secure each
individual message transmission with a constant factor overhead,
provided the message length is long enough. This can be ensured simply
by taking sufficiently long blocks. MAX belongs to the class of
type-threshold functions and can hence be computed at a rate of
$\Theta \left (\frac{1}{\log \log n}\right) $.

\section{Optimizing the time of histogram computation}
The protocol for histogram computation with binary data in a noisy
random planar network in \cite{Ying06} takes $\Theta \left (n\log \log
  n\right) $ transmissions.  Computation time is ignored. Clearly, the
time required by any correct protocol for histogram computation must
be $\Omega \left (\sqrt{\frac{n}{\log n}}\right) $ (as is true even
for MAX). We describe an algorithm that takes $\Theta \left
  (\sqrt{\frac{n}{\log n}}\right) $ time slots.

Firstly, we construct a spanning tree as in Fig.~\ref{fig:sptree}.
This ensures that the depth of the tree is limited to $\Theta \left
  (\sqrt{\frac{n}{\log n}}\right) $. This is in contrast with an
arbitrarily constructed spanning tree which can have a depth as large
as $\Theta \left (\frac{n}{\log n}\right) $. Even with the spanning
tree of Fig.~\ref{fig:sptree}, an implementation of the Inter-cell
protocol described in \cite{Ying06} requires $\Theta \left (\log
  n\right) $ time for each level of the tree, yielding a total time
requirement of $\Theta \left (\sqrt{{n}{\log n}}\right) $. Number of
transmissions needed is $\Theta \left (n\right) $. We show how to
bring the time required down to $\Theta \left (\sqrt{\frac{n}{\log
      n}}\right) $ while using the same number of transmissions.

Once again we perform Stage~2 in a series of sub-stages as shown in
Fig.~\ref{fig:substages}. Each cell needs to convey the number of
$1$'s in its sub-tree of cells (including the cell itself) to its
parent. This information can be represented by a binary string of size
$g=\lceil \log n \rceil$ for each cell (by adding appropriate zeros to
the left). Let this binary string for cell $S_j$ be $b_{g-1}^j b_{g
  -2}^j\ldots b_0^{j} ,$ where each $b^j_k$ is a bit. Consider part of
a sub-stage with the linear array of cells $S_{a_1}, S_{a_2}, \ldots,
S_{a_q}$, where $q=\Theta \left (\log n\right) $ (ignoring edge
effects). The cells are ordered with $S_{a_0}$ being the cell at the
greatest depth, and $S_{a_q}$ being the cell closest to the sink. The
noiseless protocol to be simulated using Theorem~\ref{thm:Rajagopalan}
is constructed as follows.  $C_{a_1}$ transmits $b_0^{a_1}$ up to the
cell center of its parent $C_{a_2}$ in the first time slot. Other
links have dummy bits exchanged. Before the second slot, $C_{a_2}$
computes $b_0^{a_2}$ using the number of $1$'s in cell $S_{a_2}$ and
the received bit $b_0^{a_1}$. In the second slot, $C_{a_1}$ sends
$b_1^{a_1}$ to $C_{a_2}$, and $C_{a_2}$ sends $b_0^{a_2}$ to
$C_{a_3}$. Dummy bits are sent over other links. Before the third
slot, $C_{a_2}$ computes $b_1^{a_2}$ and $C_{a_3}$ computes
$b_0^{a_3}$.  Progressing this way, at the end of $q+g -1 = \Theta
\left (\log n\right) $ slots, $C_{a_q}$ has received all required data
from $C_{a_{q-1}}$. Now, this protocol can be simulated in a noisy
setting using Theorem~\ref{thm:Rajagopalan}. The sub-stage can be
completed in $\Theta \left (\log n\right) $ time, using $\Theta \left
  (\log n\right) $ transmissions for each cell center (as for the MAX
computation). Moreover, we can ensure that the overall probability of
error goes to zero as for the MAX computation. Thus, the inter-cell
protocol can be completed in $\Theta \left (\sqrt{\frac{n}{\log
      n}}\right) $ time overall, using only $\Theta \left (n\right) $
transmissions.

\section{Conclusion}
A non-trivial lower bound is yet to be proved for number of
transmissions required for histogram computation. \cite{Goyal05}
demonstrates that $\Theta \left (n\right) $ transmissions suffice in a noisy
broadcast network (in a weak noise model). However, for a noisy random
planar network, the best known algorithm requires $\Theta \left (n\log \log
n\right) $ transmissions, whereas the trivial lower bound is $\Omega \left (n\right )$. It
would be interesting to close the gap.

Also note that the computational complexity of the algorithms that we
describe here can be significant, given that an efficient method for
decoding of tree codes is not known.

\bibliography{function-computation}

\begin{thebibliography}{10}

\bibitem{Ying06}
L.~Ying, R.~Srikant, and G.~Dullerud,
\newblock ``Distributed symmetric function computation in noisy wireless sensor
  networks with binary data'',
\newblock in {\em Proc. WiOpt}, April 2006.

\bibitem{Giridhar05}
A.~Giridhar and P.~R. Kumar,
\newblock ``Computing and communicating functions over sensor networks'',
\newblock {\em IEEE Journal on Selected Areas in Communication}, vol. 23, no.
  4, pp. 755--764, April 2005.

\bibitem{Khude05}
N.~Khude, A.~Kumar, and A.~Karnik,
\newblock ``Time and energy complexity of distributed computation in wireless
  sensor networks'',
\newblock in {\em Proceedings of the IEEE Infocom}, 2005, vol.~4, pp.
  2625--2637.

\bibitem{Gallager88}
R.~G. Gallager,
\newblock ``Finding parity in simple broadcast networks'',
\newblock {\em IEEE Transactions on Information Theory}, vol. 34, pp. 176--180,
  1988.

\bibitem{Newman04}
I.~Newman,
\newblock ``Computing in fault tolerance broadcast networks'',
\newblock in {\em 19th IEEE Annual Conference on Computational Complexity},
  2004, pp. 113--122.

\bibitem{Goyal05}
N.~Goyal, G.~Kindler, and M.~E. Saks,
\newblock ``Lower bounds for the noisy broadcast problem.'',
\newblock in {\em FOCS}, October 2005, pp. 40--52.

\bibitem{Rajagopalan94}
S.~Rajagopalan and L.~J. Schulman,
\newblock ``A coding theorem for distributed computation'',
\newblock in {\em Proc. 26th STOC}, 1994, pp. 790--799.

\bibitem{Gupta00}
P.~Gupta and P.~R. Kumar,
\newblock ``The capacity of wireless networks'',
\newblock {\em IEEE Transactions on Information Theory}, vol. 46, no. 2, pp.
  388--404, March 2000.

\bibitem{Feige00}
U.~Feige and J.~Kilian,
\newblock ``Finding or in noisy broadcast network'',
\newblock {\em Information Processing Letters}, vol. 73, no. 1-2, pp. 69--75,
  January 2000.

\bibitem{Kushilevitz98}
E.~Kushilevitz and Y.~Mansour,
\newblock ``Computation in noisy radio networks'',
\newblock in {\em SODA}, 1998, pp. 236--243.

\bibitem{Xue04}
F.~Xue and P.~Kumar,
\newblock ``The number of neighbors needed for connectivity of wireless
  networks'',
\newblock {\em Wireless Networks}, vol. 10, no. 2, pp. 169--181, March 2004.

\bibitem{Peczarski06}
M.~Peczarski,
\newblock ``Strategy in ulam's game and tree code give error-resistant
  protocols'',
\newblock {\em CoRR}, vol. cs.DC/0410043, 2004.

\end{thebibliography}
\bibliographystyle{IEEE}
\end{document}